\documentclass[usenatbib]{mn2e}    
\title[Weak lensing simulated clusters]{Mock weak lensing analysis of simulated galaxy clusters: bias and scatter in mass and concentration}
\author[Y.M. Bah\'{e} et al.]{Yannick~M.~Bah\'{e}\thanks{ybahe@ast.cam.ac.uk},$^1$ Ian G.~McCarthy$^{1,2}$ and Lindsay J.~King$^{1,2}$\\
$^1$ Institute of Astronomy, University of Cambridge, Madingley Road, Cambridge CB3 0HA, United Kingdom\\
$^2$ Kavli Institute for Cosmology Cambride, University of Cambridge, Madingley Road, Cambridge CB3 0HA, United Kingdom}
\date{6 January 2012}
\usepackage{graphicx}
\usepackage{float}
\usepackage{fancyhdr}
\usepackage{natbib}
\usepackage{amsmath}
\usepackage{amssymb}
\usepackage[usenames,dvips]{color}
\usepackage{appendix}
\usepackage{subfigure}

\newcommand{\reff}{r_\text{eff}}
\newcommand{\rlim}{r_\text{lim}}
\newcommand{\fsub}{f_\text{Sub}}
\newcommand{\mmill}{M_{\text{Mill}}}
\newcommand{\mdd}{M_{\text{3D}}}
\newcommand{\cdd}{c_{\text{3D}}}
\newcommand{\mwl}{M_{\text{WL}}}

\newcommand{\cwl}{c_{\text{WL}}}

\begin{document}
\label{firstpage}
\maketitle

\begin{abstract}
We quantify the bias and scatter in galaxy cluster masses $M_{200}$ and concentrations $c$ derived from an idealised mock weak gravitational lensing (WL) survey, and their effect on the cluster mass-concentration relation. For this, we simulate WL distortions on a population of background galaxies due to a large ($\approx 3000$) sample of galaxy cluster haloes extracted from the Millennium Simulation at $z \simeq 0.2$. This study takes into account the influence of shape noise, cluster substructure and asphericity as well as correlated large-scale structure, but not uncorrelated large-scale structure along the line of sight and observational effects (e.g., the source redshift distribution and measurement, and measurement of galaxy ellipticities). We find a small, but non-negligble, negative median bias in both mass and concentration at a level of $\sim 5\%$, the exact value depending both on cluster mass and radial survey range. Both the mass and concentration derived from WL show considerable scatter about their true values. This scatter has, even for the highest mass clusters of $M_{200} > 10^{14.8} M_\odot$, a level of $\sim 30\%$ and $\sim 20\%$ for concentration and mass respectively and increases strongly with decreasing cluster mass. For a typical survey analysing 30 galaxies per arcmin$^2$ over a radial range from $30^{\prime\prime}$ to $15^{\prime}$ from the cluster centre, the derived $M_{200}$-$c$ relation has a slope and normalisation too low compared to the underlying true (3D) relation by $\sim 40\%$ and $\sim 15\%$ respectively. The scatter and bias in mass are shown to reflect a departure at large radii of the true WL shear/matter distribution of the simulated clusters from the NFW profile adopted in modelling the mock observations. Orientation of the triaxial cluster haloes dominates the concentration scatter (except at low masses, where galaxy shape noise becomes dominant), while the bias in $c$ is mostly due to substructure within the virial radius.
\end{abstract}


\section{Introduction}
\label{sec-introduction}

There are several aspects of galaxy clusters that make them sensitive tracers of cosmic evolution, and thus potentially powerful tools for measuring a number of fundamental cosmological parameters (for recent reviews see \citealt{Voit2005} and \citealt{Allen_et_al_2011}).  First, in currently-favoured hierarchical models for structure formation where small objects collapse first and merge together to form progressively larger ones, clusters are the most recent objects to have formed, since they are the largest bound and virialized objects in the Universe at present.  Consequently, the abundance of clusters as a function of mass and redshift is sensitive to cosmological parameters that have an influence on the rate of growth of structure (such as $\sigma_8$, $\Omega_m$, and the nature of dark energy).  Second, the matter content (specifically the baryon to total mass ratio) of clusters is expected to reflect that of the universe as a whole (i.e., $\approx \Omega_b/\Omega_m$; \citealt{White_et_al_1993}), since the potential wells of clusters are incredibly deep and it is difficult to conceive of sources that are energetic enough to remove baryons from them.  Finally, the shape of the total matter density profile (which is dominated by dark matter), specifically how concentrated it is as a function of mass (e.g., \citealt{Buote_et_al_2007}), is also sensitive to the underlying cosmological parameters through the growth rate of structure, since the concentration is thought to be determined by the background density at epoch of formation (e.g., \citealt{Navarro_Frenk_White_1996}).  All of these cluster-based cosmological tests rely on there being accurate estimates of cluster mass and also concentration (in the case of the latter test, which is our main focus in the present study), at least in a statistical sense.  This therefore requires that the bias and scatter inherent in cluster mass and concentration estimators be carefully quantified and accounted for.

Weak gravitational lensing (hereafter WL) provides one of the most promising methods for deriving the masses and concentrations of large samples of galaxy clusters and thus for measuring cosmological parameters such as $\sigma_8$ and the evolution of the dark energy to (potentially) high accuracy. For high-mass clusters, masses and concentrations can be obtained individually (e.g., \citealt{Dahle_2006}; \citealt{Broadhurst_et_al_2008}; \citealt{Mahdavi_et_al_2008}; \citealt{Okabe_Umetsu_2008}; \citealt{Medezinski_et_al_2010}; \citealt{Radovich_et_al_2008}; \citealt{Corless_et_al_2009}), whereas less massive systems can be studied by stacking the lensing profiles of many clusters (e.g., \citealt{Johnston_et_al_2007}; \citealt{Mandelbaum_Seljak_2007}; \citealt{Sheldon_et_al_2009}; \citealt{Rozo_et_al_2011}). In the near future, the Dark Energy Survey (DES, \citealt{DES}) is expected to make observations of $\sim20,000$ clusters with masses in excess of $2\times 10^{14} M_{\odot}$. Note that unlike most other methods that are used to measure the mass and concentration of clusters, like hydrostatic equilibrium analyses of the hot intracluster medium (e.g.,~\citealt{Vikhlinin_et_al_2009} and references therein) or virial equilibrium analyses of orbiting satellite galaxies (e.g.,~\citealt{Becker_et_al_2007}), WL analyses do not require any assumptions about the dynamical state of the cluster.  As we will show, however, this does not guarantee that the WL-derived masses and concentrations are unbiased (see also \citealt{Becker_Kravtsov_2011} and \citealt{Oguri_Hamana_2011}). 

In the present study we aim to quantify the bias and scatter in galaxy cluster masses, $M_{200}$, and concentrations, $c$, derived from a mock WL survey and their effect on the derived cluster mass-concentration relation.  Mock WL observations based on numerical simulations and analytic models of clusters offer the unique possibility of comparing the observed and true parameters for the \emph{same} clusters to find the extent to which observational biases contribute to the discrepancies.  There have been a number of such studies in the past, focussing on various aspects such as cluster asphericity and substructure (e.g.,~\citealt{King_et_al_2001}; \citealt{Clowe_et_al_2004}; \citealt{Corless_King_2007}), correlated (e.g.,~\citealt{Metzler_et_al_1999}; \citealt{Metzler_et_al_2001}; \citealt{King_Corless_2007}) and uncorrelated large-scale structure along the line of sight (e.g., \citealt{Hoekstra_2003,Dodelson_2004,Hoekstra_et_al_2011}). In particular, \citet{Corless_King_2007} found from an analysis of lensed analytic clusters that for projections along a line of sight close to the major axis concentrations and mass can be overpredicted by factors of up to $\approx 2$ and 1.5 respectively. In a comparison between X-ray and WL derived masses of galaxy clusters, \citet{Meneghetti_et_al_2010} used high-resolution re-simulations of three clusters to find deviations of up to 50\% between true and WL masses.
 
While these studies indicated the extent to which various complications \emph{can} potentially bias the WL-derived masses and concentrations, it is equally important to know \emph{how likely} they will do so.  Investigating this requires large, high resolution cosmological simulations containing a large enough number of realistically modelled clusters to allow statistically robust and meaningful conclusions to be drawn about those in the real Universe.  In a recent study aiming at this, \citet{Becker_Kravtsov_2011} --- see also \citealt{Oguri_Hamana_2011} --- used a large sample of $\sim 10^4$ haloes and derived a typical scatter in the reconstructed \emph{masses} of $\sim 25 - 30$\% and also a systematic bias of $\approx -5$ to $-10\%$.  In the context of the mass-concentration relationship however, which is our focus here, we also need to quantify the corresponding scatter and bias in concentration, as well as any potential variation of it with cluster mass.   
 
With this aim in mind, we produce synthetic WL observations based on a large, unbiased sample of galaxy clusters from the Millennium Simulation, each of which is `observed' in a number of different, randomly selected projections. These mock observations are then analysed in approximately the same way as observational data and the mass and concentration obtained are compared to the true values to quantify the scatter and bias in the reconstructed parameters and the mass-concentration relation.  Beyond quantifying the scatter and bias, we also aim to identify physical explanations for them from a detailed analysis of the simulated clusters. This may help to  miniminse these effects in future surveys, for example through appropriate selection of clusters or limiting the analysis to a particular radial range.  
 
The simulations we use contain only dissipationless dark matter.  While including a realistic baryonic component would be desirable, the computational cost of running hydrodynamic simulations of large volumes (a requirement for mock WL surveys) at high resolution is currently prohibitively large. While baryonic physics is expected to significantly impact on the \emph{strong} lensing properties of massive clusters (e.g., \citealt{Mead_et_al_2010}), it has recently been shown by \citet{Semboloni_et_al_2011} that its effect on WL measurements is limited to haloes with a mass below $10^{14} M_\odot$. Analysing exclusively clusters above this mass threshold, we therefore do not expect the inclusion of a realistic baryonic component to significantly modify our results or conclusions.

This paper is structured as follows: In \S2 we review the basic WL formalism and then describe our simulations and subsequent data analysis in \S3. We quantify the mass-concentration relation and the spread in the reconstructed parameters in \S4, followed by a detailed analysis of the physical principles responsible for scatter and bias in \S5. We discuss and summarise our findings in \S6.  For consistency with the Millennium Simulation, a flat cosmology with Hubble parameter $h =$ H$_{0}/(100\,{\rm km}\,{\rm s}^{-1}{\rm Mpc}^{-1}) = 0.73$, dark energy density parameter $\Omega_\Lambda = 0.75$ (dark energy equation of state parameter $w=-1$), and matter density parameter $\Omega_{\rm M} = 0.25$ is used throughout this paper.

\section{Weak Lensing Formalism}
\label{sec-lensingtheory}
Gravitational lensing describes the deflection of light by massive objects. It is sensitive only to the projected mass density $\Sigma$ of the lens, with higher $\Sigma$ generally corresponding to larger distortions, for centrally condensed objects. Unlike strong lensing, which occurs in cluster centres and leads to 
multiple images of background galaxies which may be highly distorted (eg. \citealt{Soucail_et_al_1987a}), weak gravitational lensing can be employed to study the outer ($r \ga 100$ kpc), less dense cluster regions.  Due to its weak nature, background galaxies are only slightly distorted, an effect that must be studied statistically. Assuming that the light from galaxies can be approximated by elliptical isophotes, one can quantify their shape by the complex quantity ellipticity ($\epsilon$), its modulus related to the ratio between the minor and major axis $r=b/a$ by 
\begin{equation*}
|\epsilon| = \frac{1-r}{1+r}
\end{equation*}
\noindent and with a phase of twice the position angle of the ellipse's major axis. In the WL regime, the observed and source ellipticities of a background galaxy, in the following denoted by $\epsilon$ and $\epsilon_s$ respectively, are related by
\begin{equation}
\epsilon = \frac{\epsilon_s + g}{1 + g^*\epsilon_s}
\label{eq-distortion}
\end{equation} 
\noindent where the complex quantity $g$ is the ``reduced shear'', with its complex conjugate denoted by $g^*$. The reduced shear is then related to the (complex) shear, $\gamma$ (the tidal gravitational field), and the convergence, $\kappa$, by 
\begin{equation}
g = \frac{\gamma}{1-\kappa}\,.
\label{eq-g}
\end{equation}
Equation \eqref{eq-distortion} implies that in the ideal case of a perfectly circular background galaxy with r = 1 and $\epsilon_s = 0$, the lensed ellipticity $\epsilon = g$. 

In the above, $\kappa$ is proportional to the projected mass density of the lens, $\Sigma$:
\begin{equation}
\kappa = \frac{\Sigma}{\Sigma_{\text{cr}}}\,,
\label{eq-kappa}
\end{equation}
\noindent where
\begin{equation}
\Sigma_{\text{cr}} = \frac{c^2}{4\pi G}\frac{D_s}{D_d D_{\mathrm{ds}}}
\label{eq-sigcrit}
\end{equation}
\noindent is the ``critical density'' and $D_s$, $D_d$ and $D_{ds}$ are the angular diameter distances between observer -- source, observer -- lens and lens -- source respectively. In this notation, the WL regime is where $\Sigma << \Sigma_{\text{cr}}$, i.e. $\kappa <<1$. 

The lensing deflection potential, $\psi$, is related to $\gamma$ and $\kappa$ by a set of partial differential equations (e.g. \citealt{Bartelmann_Schneider_2001}):
\begin{eqnarray}
\kappa(\mathbf{\theta}) & = & \frac{1}{2}\left(\psi_{,11} + \psi_{,22}\right) \label{eq-gamma} \\
\gamma_1(\mathbf{\theta}) & = & \frac{1}{2}\left(\psi_{,11} - \psi_{,22}\right) \nonumber \\
\gamma_2(\mathbf{\theta}) & = & \psi_{,12} = \psi_{,21} \nonumber\,,
\end{eqnarray}
where
\begin{eqnarray}
\gamma(\mathbf{\theta}) & = & \gamma_1 + i\gamma_2 \nonumber
\end{eqnarray}
\noindent and the indices following the comma denote partial derivatives with respect to the components of the position vector $\mathbf{\theta}$.

A secondary effect is a slight modification in the observed number density of background galaxies, $n_{\rm lensed}$, due to magnification changing the apparent fluxes of galaxies, as well as changing the apparent area of sky in which they are observed. The net effect depends on the slope of the number counts of the galaxies ($\beta$), with the lensed and unlensed number counts being related by \citep{Canizares_1982}:
\begin{equation}
n_{\text{lensed}} = \mu^{\beta-1}n_{\text{unlensed}}
\label{eq-numbercount}
\end{equation}
where the magnification $\mu$ is given by 
\begin{equation}
\mu = \frac{1}{(1-\kappa)^2-|\gamma|^2}\,.
\label{eq-mu}
\end{equation}
Since $\beta\approx 0.5$ for the faint distant galaxies typically used in WL analyses, this results in ``number count depletion" \citep{Broadhurst_et_al_1995}, with the observed background galaxy density decreasing with decreasing distance from the cluster centre in the WL regime. 

\section{Mock Weak Lensing Observations and Analysis}
\label{sec-simulations}

\subsection{Simulated galaxy clusters}

We extract simulated galaxy clusters from the Millennium Simulation (hereafter MS; \citealt{Springel_et_al_2005}), a large cosmological N-body simulation that follows $2160^3$ dark matter particles from $z=127$ to $z=0$ in a periodic box of $500\,h^{-1}$\,Mpc on a side. The cosmology adopted for the MS, which we also adopt for our analysis, is a flat $\Lambda$CDM model with $h=0.73$, $\Omega_{\rm M} =0.25$, and a power spectrum normalisation on a scale of $8 h^{-1}$Mpc of $\sigma_8 = 0.9$ (i.e., the rms linear mass fluctuation in a sphere of radius $8 h^{-1}$ Mpc extrapolated to present-day). These parameters are consistent with the latest measurements of temperature and polarization anisotropies in the cosmic microwave background (CMB) with the \textit{Wilkinson Microwave Anisotropy Probe} (WMAP; \citealt{Komatsu_et_al_2011}), although the value of $\sigma_8$ adopted for the MS is larger than the maximum likelihood CMB value by $\approx 2\sigma$. The larger value of $\sigma_8$ means that the MS will have more massive clusters than a universe with the WMAP 7-year cosmology. However, as we are interested in the lensing signal of \emph{individual} clusters (specifically how well WL can recover their mass and concentration) and not their abundances, this discrepancy will not affect the validity of our results.
 
From the snapshot at $z \simeq 0.2$, a typical redshift of observed WL clusters, we select all of the simulated clusters with masses $M_{200} \ge 10^{14} M_\odot$ for analysis. Note that $M_{200}$ is defined as the mass within a radius ($r_{200}$) that encloses a mean density equal to 200 times the critical density of the universe --- in our adopted cosmology, this density is $\rho_{\text{crit}} = 1.76 \cdot 10^{11} M_\odot / \text{Mpc}^3$ at our chosen redshift. From here on, we will refer to these masses obtained directly from the MS particle distribution without any profile fitting as `true' mass, denoted by $\mmill$.  A large sample of 2678 simulated cluster haloes satisfies this criterion, which will allow us to robustly quantify the mean trends and scatter in the derived mass-concentration relationship. 

We extract all of the dark matter particles within a $10\,h^{-1}$\,Mpc (comoving) box, centered on the most bound particle, for producing WL maps of each cluster. For reference, $r_{200}$ is typically $\approx 2$ Mpc for the most massive clusters in our sample (and smaller for lower mass clusters). Thus, our analysis includes only the {\it local} environment around the clusters. This was deliberate, so that we can isolate the effects of cluster triaxiality, substructure and connecting filaments (i.e., ``correlated'' signals) from uncorrelated line-of-sight structures at much larger distances. While these correlated structures can in general extend beyond a distance of 5 $h^{-1}$ from the cluster centre, we verify in appendix \ref{sec:testlength} (Fig.~\ref{fig:boxsizetest}) that, in agreement with the findings by \citet{Becker_Kravtsov_2011}, our results are robust to increasing the line-of-sight integration length by a factor of five to $50\,h^{-1}$\, comoving Mpc. Much longer integration lengths, which would capture uncorrelated large scale structure more fully, would in principle be desirable for the purpose of improving the prediction accuracy, but require the use of ray-tracing algorithms which is beyond the scope of this work. In a complementary study, \citet{Hoekstra_et_al_2011} have used ray-tracing of the MS to examine the effect of such uncorrelated structures on the derived masses and concentrations of analytic clusters and found that it contributes to scatter, but not bias in mass and concentration. We discuss the results of these authors in the context of our findings in Section \ref{sec:conclusion}.

\begin{figure*}
\includegraphics{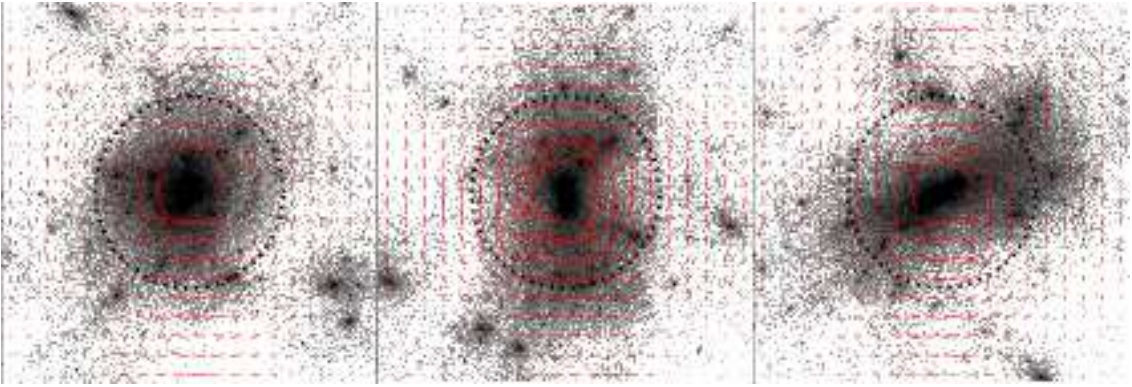}
\caption{Three projections of a high-mass cluster with M$_{200} = 4 \cdot 10^{15} M_\odot$. The red tick marks represent the reduced shear field induced by these projections in the WL regime (shown for $\kappa < 0.1$), with the tick length proportional to the shear modulus and orientation showing the shear angle. The scaling is logarithmic, darker regions have higher density. The circles enclose the region of $15^\prime$ around the cluster centre which was analysed in this study.  Different projections of the same cluster can give rise to very different shear signals.}
\label{projections}
\end{figure*}

\subsection{Generating weak lensing maps}

For each cluster we produce surface mass density maps by projecting the $10\,h^{-1}$\,Mpc box along one side and interpolating to a regular grid of 2000$^2$ pixels using a triangular shaped clouds (TSC) algorithm \citep{Hockney_Eastwood_1988}. It was checked that other interpolation algorithms, such as cloud-in-cells or smoothed particle hydrodynamics (SPH) interpolation, did not yield significantly different results. At the adopted resolution, 1 pixel length corresponds to $5\,h^{-1}$ kpc, equivalent to the gravitational softening length of the MS, which makes it the highest sensible resolution. 

To further boost our statistics, and explore the role of viewing angle for individual objects, we repeated this procedure for each simulated cluster five times, each time selecting a random orientation and thus produce five different maps for each cluster. In the DES, observations will be made of $\sim20,000$ clusters with masses in excess of $2\times 10^{14} M_{\odot}$. The total number of projections we produce is comparable to this survey.
For the most massive clusters (with $M_{200} \ge 10^{14.8} M_\odot$) in our sample, we have elected to boost our statistics even further by producing maps corresponding to 50 different random projections. The logic behind this is that i) such massive systems are the most commonly-targetted in present WL observational studies (and therefore are the systems for which we can presently make comparisons between theory and observations); and ii) very high mass ($\sim 10^{15} M_\odot$) clusters are rare even in a $500 h^{-1}$ Mpc volume.

Convergence ($\kappa$) maps are then computed according to eqn.\,\eqref{eq-kappa} with $\Sigma_{\rm cr}$ as above, and the shear maps formed according to equations \eqref{eq-gamma} and \eqref{eq-g}. To solve these equations, we adopt the same FFT method as \citet{Clowe_et_al_2004}, computing
\begin{equation}
\tilde{\gamma} = \left(\frac{\hat{k}_1^2-\hat{k}_2^2}{\hat{k}_1^2+\hat{k}_2^2}\tilde{\kappa}, \frac{2\hat{k}_1\hat{k}_2}{\hat{k}_1^2+\hat{k}_2^2}\tilde{\kappa}\right)
\label{eq:fftmethod}
\end{equation}
where $\tilde{\gamma}$ and $\tilde{\kappa}$ are Fourier transforms of $\gamma$ and $\kappa$, and $\hat{k}$ are the appropriate wave vectors. As an example, Fig.~\ref{projections} shows three projections of one high-mass cluster with the corresponding shear fields as tick marks. It is apparent that different projections of the same cluster can look very different and lead to different shear signals.

With shear maps for all the projections generated, an array of background galaxies, extending by default to a projected distance of $r_\text{out} = 15^\prime$ from the cluster centre, was then simulated for each individual projection. This limit in projected distance excludes potential numerical artifacts near the edges of the shear field; it also matches the typical field-of-view of observations (e.g. \citealt{Clowe_Schneider_2001}; see also \citealt{Hoekstra_2003}), where for the most massive clusters the shear signal is detected out to roughly this radius. Although galaxies used in real weak lensing surveys have a range of redshifts, we consider only the simplified case of a uniform source redshift of $z = 1.0$\footnote{Note that at the lens redshift of $z \simeq 0.2$, $1^\prime$ corresponds to $\sim 0.2$ Mpc for our adopted cosmology.}. This is a value typical for observational studies such as CFHTLS \citep{Ilbert_et_al_2006} and COSMOS \citep{Ilbert_et_al_2009} and has also been used in other theoretical work concerning the accuracy of weak lensing (e.g., \citealt{Clowe_et_al_2004}; \citealt{Oguri_Hamana_2011}; \citealt{Gruen_et_al_2011}; \citealt{Becker_Kravtsov_2011}). In real surveys the dispersion in source redshifts and its finite sampling \citep{Hoekstra_et_al_2011} has to be considered as an additional error source. We note that for this choice of lens and source redshift, the critical surface mass density $\Sigma_\text{cr} = 3.33 \cdot 10^{15} M_\odot / $Mpc$^2$.

To exclude the strong lensing regime in the cluster centre, we excised the central region within $r = r_\text{in}$, with a default inner limit radius $r_\text{in} = 30$ arcsec, a value typical of the extent of strongly lensed arcs in observational studies (e.g., \citealt{Hennawi_et_al_2008}). We consider the effect of changing the values of both $r_\text{in}$ and $r_\text{out}$ in appendix \ref{sec:cuttest}; while the exact amount of bias and scatter depends somewhat on the exact choice of $r_\text{in}$ and $r_\text{out}$ (see also \citealt{Becker_Kravtsov_2011}) our overall conclusions are unaffected by their exact choice. 

The background galaxy ellipticities $\epsilon_s$ were drawn randomly from a Gaussian distribution with standard deviation $\sigma = 0.2$ per component as in \citet{Geiger1998}; note that this choice of $\sigma$ is also similar to that used by \citet{Hoekstra_et_al_2011}. The effect of increasing the value of $\sigma$ is shown in appendix \ref{sec:sigmatest}; it does not affect our results significantly. The positions of galaxies were allocated randomly, with average unlensed density from which shapes can be measured accurately $n = 30$ arcmin$^{-2}$, and accounting for Poisson noise. This is comparable to background densities achievable with current ground-based pointed observations of galaxy clusters (e.g., \citealt{Dahle_2006}) and similar to what will be achieved with upcoming large surveys such as the DES\footnote{Even though the expected source density for the DES, of $\sim 10$ arcmin$^{-2}$, is somewhat lower than our adopted fiducal background density, we have experimented with lowering our source density from 30 to 10 arcmin$^{-2}$ and find that it results in only a modest increase in the scatter in the derived masses and concentrations. Our conclusions in Section 5, that it is halo triaxiality and substructure, rather than low signal-to-noise, that dominate the error in the reconstructed masses and concentrations are robust to lowering the background source density to this level.} and that from the Large Synoptic Survey Telescope (LSST)\footnote{http://www.lsst.org/lsst/}. 

The distortion of these simulated background galaxies by the cluster-induced shear field was then computed according to eqn.\,\eqref{eq-distortion} to yield the lensed background galaxy ellipticities $\epsilon$. To account for number count depletion as described in Section \ref{sec-lensingtheory}, the lensed source density was adjusted locally according to equations \eqref{eq-numbercount} and \eqref{eq-mu}. At this stage we have the final synthetic catalogues of lensed galaxies.

\subsection{Fitting cluster profiles}
\label{sec:fitting}

We derive the best-fit mass, $M_{200}$, and concentration, $c$, by fitting projected NFW mass distributions  (\citealt{Navarro_Frenk_White_1995}; \citealt{Navarro_Frenk_White_1996}; \citealt{Navarro_Frenk_White_1997}) to our mock WL data. The NFW profile, which has been shown to reproduce the 3D spherically-averaged mass density profiles of dark matter haloes spanning a wide range of masses, is described by the two parameter form:

\begin{equation}\label{nfwrho}
\rho (r) = \frac{\rho_s}{(r/r_s)(1+r/r_s)^2} \,,
\end{equation}

\noindent where

\begin{equation}
\rho_s/\rho_{\rm crit} = \frac{200}{3}\frac{c^3}{\ln(1+c)-c/(1+c)}
\end{equation}

\noindent and the concentration $c \equiv r_{200}/r_s$. Thus, the density distribution is determined once the concentration (or equivalently the scale radius, $r_s$) and $M_{200}$ (or equivalently $r_{200}$) are specified.

Analytic expressions for the NFW profile integrated along the line of sight, $\Sigma_{\text{NFW}}$, and the resulting shear $\gamma$ are given by \citet{Bartelmann_1996} and \citet{Wright_Brainerd_2000} respectively, from which the WL distortion can be calculated as shown above. We point out that the analytic expressions found in these studies were derived by integrating an NFW profile extending to \emph{infinity} along the line of sight, rather than to some distance characteristic of the cluster size (e.g., $r_{200}$). We discuss the effect of this model assumption further in Section \ref{sec:mass}. 

We derive the best-fit NFW parameters by first calculating the tangential ellipticity $\epsilon_t$ of all background galaxies with cluster-centric distance $r_\text{in} < r < r_\text{out}$, defined by 
\begin{equation}
\epsilon_t = -\text{Re}\left\{\epsilon e^{-2i\phi}\right\}
\end{equation}
\noindent which are then fit by the tangential reduced shear generated by an NFW profile, parameterised by $M_{200}$ and $c$, using a standard least-squares method (see \citealt{Press_et_al_1992}) with metric 
\begin{equation}
\label{chisq}
\chi^2 = \displaystyle\sum\limits_{i=0}^N \left[\epsilon_{t,i}-g_{\text{NFW}}(r_i, M,c)\right]^2.
\end{equation}
Equation \eqref{chisq} implies that we take into account, equally weighted, the tangential ellipticity of each individual galaxy when fitting. Other commonly used methods employed to analyse weak lensing signals are maximum-likelihood techniques (e.g.,~\citealt{Corless_King_2008}) and fitting a 1D shear profile derived from binning up the individual galaxy ellipticities (e.g.,~\citealt{Clowe_et_al_2004}). We have explicitly verified that our particular choice of fitting method has no significant influence on our results (see also appendix \ref{sec:cuttest}). The best-fit concentration and mass are then taken as the reconstructed cluster parameters. From here on, we will refer to the values obtained in this way as $c_{\text{WL}}$ and $M_{\text{WL}}$ respectively.

Our aim is to compare WL reconstructions to the real cluster parameters to see how projection effects and the application of a simplified mass model for the halo modify the deduced values. For this comparison it is also necessary to obtain a `true' concentration --- a model-independent reference value for the mass is already provided by $\mmill$ --- which we do by fitting an NFW profile to the 3D cluster data. In the following, we will denote the parameters M$_{200}$ and $c$ derived in this way by $\cdd$ and $\mdd$. 

Our 3D fitting procedure is similar to that described by \citet{Neto_et_al_2007} for haloes in the Millennium simulation: the cluster particles are grouped into 32 radial bins from $\log_{\text{10}} (r/r_{\text{200}}) = -2.5$ to 0.0, equally spaced in $\log_{\text{10}} (r/r_{\text{200}})$. For each bin, the average density $\rho_i$ is then computed and the NFW parameters obtained by performing a least squares fit of $\log_{\text{10}} \rho_i$ to $\log_{\text{10}} \rho_{\text{NFW}}$, taking into account only bins at a radius $\log_{\text{10}} (r_i/r_{200}) \geq -1.63$. This criterion ensures that the central cluster region, where the density is affected by smoothing, is excluded from our fit. In this, our method differs slightly from \citet{Neto_et_al_2007}, who adopt an inner cut-off radius of $\log_{\text{10}} (r_i/r_\text{vir}) = -1.3$. The reason behind this is that unlike their study, we only look at massive haloes with $M_{200} \geq 10^{14}$ M$_\odot$ with a larger $r_{200}$. It has been shown by \citet{Gao_et_al_2008} that the best-fit concentration in the NFW profile depends somewhat on the exact choice of the inner cut-off radius, with lower values generally leading to higher concentrations, which we have confirmed for our cluster fits. It should therefore be borne in mind that the bias we derive is strictly applicable only with respect to this particular fitting range. In the future, it may be advisable to adopt more accurate fitting functions such as the Einasto profile \citep{Einasto_1965} as discussed by \citet{Navarro_et_al_2004} and \citet{Merritt_et_al_2006}; however, the price to pay for this improved accuracy is the introduction of an additional degree of freedom and its potential correlations with $M_{200}$ and $c$. 

We point out in passing that we have not imposed any relaxation criteria for selecting our simulated clusters and select based on mass alone, as it is not trivial to deduce how relaxed a real cluster is. It is known that the NFW profile does not describe obviously unrelaxed cluster haloes well (e.g., \citealt{Neto_et_al_2007}) and thus we expect that the accuracy of the reconstructed mass and concentration of real systems will depend on the dynamical state of the cluster.

\begin{figure}
\includegraphics[width=\columnwidth]{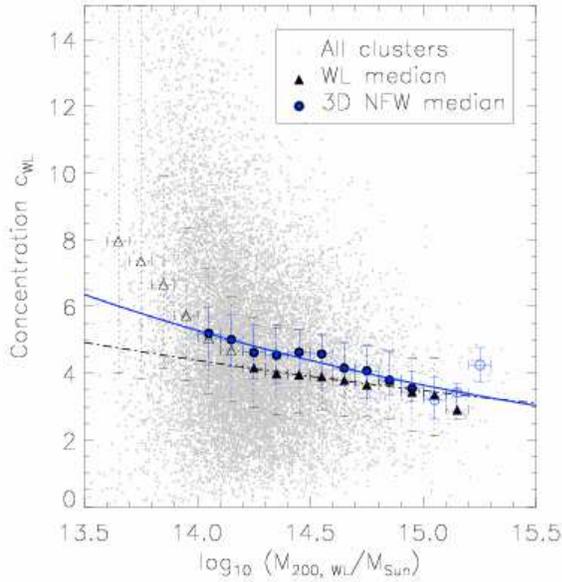}
\caption{Mass-concentration plot obtained from our mock weak lensing analysis showing five projections of each cluster (grey points). The black triangles show the median concentrations for each bin with the black dash-dot line giving the corresponding best-fit power-law as described in the text. The blue circles and solid line give the corresponding 3D NFW medians and best-fit power-law respectively. Filled symbols represent median values used for constructing the power-law, open ones were discarded. The errorbars (dotted for WL, solid for 3D) indicate binsize in x-direction, whereas in y-direction the 25$^\text{th}$ and 75$^\text{th}$ percentiles are shown. The weak-lensing derived concentrations are systematically too low compared to those from our reference 3D fit, with the discrepancy increasing for lower mass systems.}
\label{mcplot}
\end{figure}
 
\section{Derived mass-concentration relation}

In Fig.~\ref{mcplot} we show the mass-concentration relationship derived from our mock WL analysis of MS clusters. Each of the grey dots represents a single projection of one cluster. The solid triangles represent the median WL-derived concentration in 16 equally-spaced bins of $\log_{10} (\mwl/M_\odot$) from 13.6 to 15.2 with bin width $\Delta\log_{10} (\mwl/M_\odot) = 0.1$. The solid circles represent the median true (3D) concentration in the corresponding true mass bins . 

\begin{figure*}
\includegraphics[width=160mm]{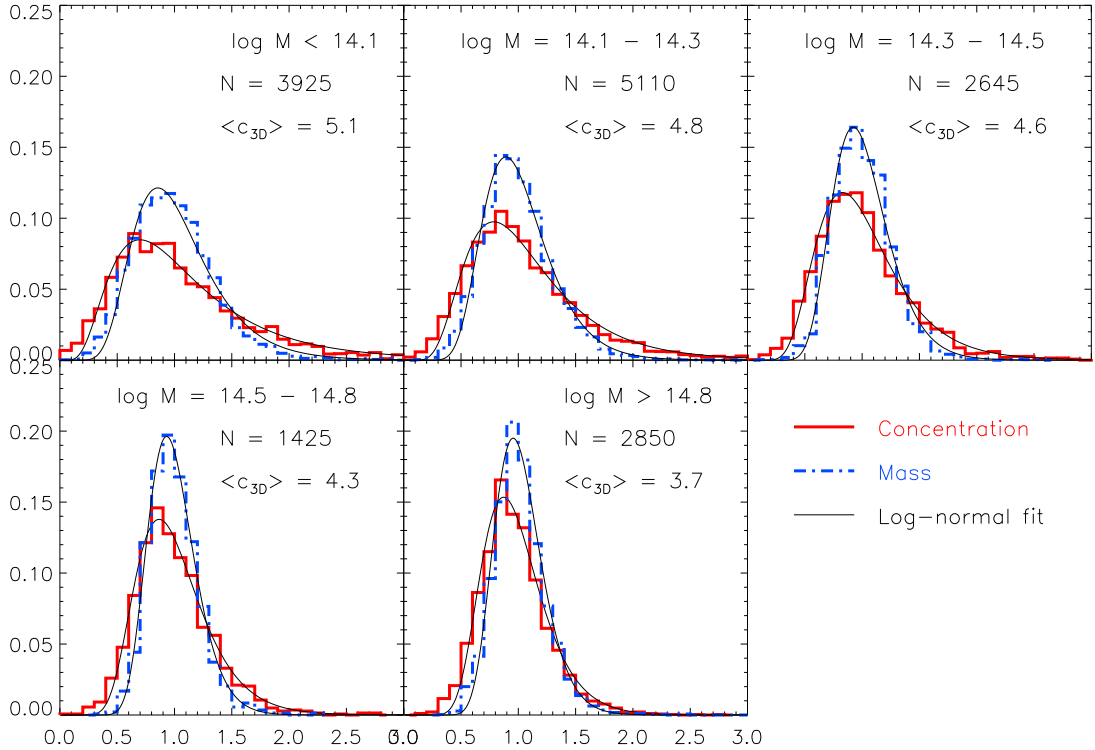}
\caption{Spread in WL concentrations $c_\text{WL}/c_\text{3D}$ (red, solid line) and masses $M_\text{WL}/M_\text{Mill}$ (blue, dash-dot line) for different mass ranges. Shown is the fraction of projections in each bin. The solid black lines show the corresponding best-fit log-normal distributions as described in the text, with parameters given in table \ref{table}. The numbers in the figure give the average concentration, mass range and number of projections per bin.  The scatter in the recovered masses and concentrations decreases with increasing halo mass and closely follow log-normal distributions.} 
\label{histogram}
\end{figure*}

\begin{table*}
\begin{tabular}{l|ccccc|ccccc} \hline
\multicolumn{1}{c}{} & \multicolumn{5}{c|}{Concentration} & \multicolumn{5}{c}{Mass} \\ \hline
$\log_{\text{10}} (M_{200}/M_\odot)$ & $\mu$ & $\sigma$ & $<c/c_{3D}>$ & median & $Q_{0.975}$ & $\mu$ & $\sigma$ & $<M/\mmill>$ & median & $Q_{0.975}$\\ \hline
14.0 -- 14.1 &  -0.033 & 0.571 &  1.163 & 0.947 & 3.58 & -0.026 & 0.362 & 0.999 & 0.957 & 1.82 \\
14.1 -- 14.3 &  -0.023 & 0.467 & 1.083 & 0.961 & 2.55 & -0.026 & 0.299 & 0.996 & 0.965 & 1.68 \\
14.3 -- 14.5 &  -0.045 & 0.380 & 1.010 & 0.946 & 1.99 & -0.016 & 0.255 & 1.000 & 0.980 & 1.54 \\
14.5 -- 14.8 & -0.043 & 0.318 & 0.989 & 0.942 & 1.74  & -0.026 & 0.213 & 0.992 & 0.968 & 1.47 \\
$>$ 14.8 & -0.056 & 0.286 & 0.961 & 0.932 & 1.63 & -0.002 & 0.209 & 1.014 & 0.992 & 1.52 \\ \hline
\end{tabular}
\caption{Parameters of the log-normal fit to the distributions of mass and concentration deviations as described in the text. Also given are the mean, median and $Q_{0.975}$ quantile for both distributions. Note that $\mu$ and $\sigma$, the median and standard deviation in the underlying normal distribution, are the location and spread in the log-normal fit as defined in equation \eqref{lognormalfit}.}
\label{table}
\end{table*}

There is considerable scatter in the mass-concentration relation derived from the mock WL observations, but the trend towards higher concentrations for lower masses is evident. A power-law of form 
\begin{equation}
\alpha (M_{\text{WL}}/10^{14} h^{-1} M_\odot)^{\beta}
\end{equation}
\noindent was fit to the WL-derived median mass-concentration distribution, assuming an error in each bin proportional to $1/\sqrt{n_i}$ where $n_i$ is the number of cluster projections per mass bin. Only mass bins with $\mwl > 14.2$ were included in the fit, since below this mass range the results are affected by our arbitrary cut-off at $\mmill = 10^{14} M_\odot$. The best-fit parameters obtained from this procedure are $\alpha_\text{WL} = (4.25 \pm 0.04)$ and $\beta_\text{WL} = (-0.10 \pm 0.01)$; we show this relation as the solid black line in Fig.~\ref{mcplot}. For comparison, we also show the best-fit powerlaw to the true (3D) mass-concentration relation (solid blue line in Fig. \ref{mcplot}), determined in the same way as for the WL data, but excluding systems with $\mdd > 10^{15} M_\odot$ due to the very small number of systems in this range. The parameters of this best-fit 3D powerlaw are $\alpha_{3D} = (5.02 \pm 0.08)$ and $\beta_{3D} = (-0.16 \pm 0.02)$. Both the normalisation and slope of the WL inferred powerlaw are too low compared to their 3D counterparts. 

\subsection{Quantifying the spread in the masses and concentrations}
\label{sec:spread}
The results in the previous section underline the need to quantify and account for bias and scatter in observationally derived cluster masses and concentrations in large surveys. For detailed studies of individual clusters the bias may be less relevant, but knowing the expected scatter is still important. In this section, we present a quantification of the bias and scatter in our mock WL derived masses and concentrations. As variation with halo mass and resulting strength of the lensing signal can be expected, our sample was first divided into five (true) mass bins, as indicated in Table \ref{table}. For each of these five bins, a histogram of the relative masses and concentrations, normalised to true mass $\mmill$ and best-fit 3D concentration $c_\text{3D}$, was then created. The results are shown in Fig.~\ref{histogram}. 

The spread in both $M_\text{WL}$ and $c_\text{WL}$ is clearly mass-dependent and decreases with increasing cluster mass. Over-concentrations of more than a factor of 2.5 are virtually non-existent except for the lowest mass bin where there is an, albeit small, group of cluster projections whose concentrations are overestimated by up to a factor of 3. The masses are somewhat better constrained, over- and underpredictions by more than a factor of 2 being rare in all mass bins (apart from the very lowest). 

The error distributions in mass and concentration were fit with a log-normal distribution
\begin{equation}
\label{lognormalfit}
f(x) = \frac{0.1}{x\sqrt{2\pi\sigma^2}} e^{-\frac{(\ln x-\mu)^2}{2\sigma^2}}
\end{equation} 
where the prefactor of 0.1, equal to the bin size in Fig.~\ref{histogram}, converts the probability distribution function into the relative histogram density shown in this figure. As shown in Fig.~\ref{histogram}, the best fit provides an excellent representation of the error distributions. The corresponding parameters are quoted in Table \ref{table} together with the median and mean obtained directly from the error distributions. Also given are the $Q_{0.975}$-quantiles, which give the relative error exceeded by only 2.5\% of projections as an indication for ``reasonably'' likely overconcentrations expected in observations. 

We point out here that, dealing with non-Gaussian distributions, the conventional (arithmetic) standard deviation is only of limited applicability. For the remainder of this paper, we will therefore use $\sigma$ for the spread parameter (the standard deviation in the underlying normal distribution) as defined in equation \eqref{lognormalfit} and `scatter' for the offset from unity in the geometric standard deviation, $e^\sigma-1$. 

Apart from large scatter, both the concentration and the mass also display a slight bias in the sense that, on average, the reconstructed values are slightly lower than the true ones. This accounts for the lower-than-expected normalisation of the WL-derived mass-concentration relation evident in Fig.~\ref{mcplot}. We aim to provide an explanation for these biases in Section \ref{sec:errorsources} below. 

Note that we have chosen to use medians, rather than means to quantify bias. The rationale behind this is that, in a non-Gaussian distribution as is the case here (log-normal), the mean, unlike the median, depends on the scatter as overpredictions can be arbitrarily high, whereas values can clearly not be underpredicted by more than 100\%. From the amount of scatter evident in our results (see Figs.~\ref{histogram} and \ref{fig:scatter}) one should not be surprised to find that the mean mass and concentration show, in general, a positive bias with respect to their median counterparts. For completeness, we show the difference between mean and median bias in Fig.~\ref{fig:mean} in appendix \ref{sec:mean}.

\subsection{Comparison to previous work}
\subsubsection{Becker \& Kravtsov (2011)}
The bias and scatter in cluster masses derived from weak lensing were recently studied in detail by \citet[BK11]{Becker_Kravtsov_2011} using clusters formed in a cosmological simulation independent of the MS run at somewhat lower resolution. In this study, the mass derived from WL was found to be biased low at a level of $\sim -5\%$. Considering that these authors employed a slightly different reconstruction method in which background galaxies were used over a radial range from 1$^\prime$ to 20$^\prime$ from the cluster centre and a shear profile formed from them, our results are in good quantitative agreement (see also Fig.~\ref{fig:cuttest} in the appendix, where we analyse our simulation using the same radial range as BK11). The scatter determined by BK11 ($\sim 20\%$) is very close to the level we derive ($\sim 25\%$); we note, however, that these two numbers were derived using two slightly different analysis methods. 

\subsubsection{Oguri \& Hamana (2011)}
\label{sec:oguriandhamana}
Biases in both mass and concentration have also been studied by \citet[OH11]{Oguri_Hamana_2011}, who found a mass bias similar to that in \citet{Becker_Kravtsov_2011} and presented here.  However, in contrast to our results, they find a very large, positive concentration bias of $\sim 20\%$.  It is presently unclear what the origin of this difference is.  We speculate that it may originate from a difference in the weak lensing simulation method between our two studies: while our results are based on direct fitting of a high resolution weak lensing simulation, OH11 analyse an analytic circularly symmetric shear profile that was derived by stacking (mock) ray-traced lensing observations of galaxy clusters formed in a low resolution cosmological simulation.  The azimuthal averaging is expected to smooth out the presence of substructure and triaxiality, both of which tend to bias the concentration low (see \ref{sec:biasorigin}). 

\section{Origin of scatter and bias}
\label{sec:errorsources}
Having established the extent of the scatter and bias in WL reconstructions of cluster haloes, we now aim to find physical explanations for them. This is an interesting question in its own right, but might also allow an identification of possible ways to reduce these systematic errors.

Any potential error sources can be broadly grouped into two categories: Those due to the background galaxies used in the reconstruction (i.e., their unknown intrinsic ellipticities and finite number, in general also their intrinsic alignment due to cosmic shear), and those due to the cluster itself, such as halo triaxiality and substructure. In this section we show that, in the case of large statistical samples of clusters such as we have studied here and those to be derived from the DES and LSST, the latter is dominated by the former only for clusters with masses below a few $10^{14} {\rm M}_\odot$.

Our strategy for assessing the importance of these various error contributions involves making two additional reconstructions of our cluster sample, designed to bridge the gap between the WL analysis based on particles within a 10 $h^{-1}$ Mpc box on the one side and the 3D fitting procedure within a radius $r_{200}$ on the other. In the first of these, which we will refer to as ``perfect WL'', we use a very high density ($n = 300$ arcmin$^{-2}$) of perfectly circular background galaxies (i.e., $\sigma = 0.0$), which eliminates the influence of shape noise and sampling\footnote{We have explicitly verified that there is no significant difference between $n = 100$ and $n = 300$ for $\sigma = 0.0$.} --- essentially, we are now analysing the (reduced) shear field $g$ directly. 

In the second method, we approach the 3D fit even further by constructing the convergence field $\kappa$ --- and thus the shear --- only from those particles that lie within a (3D) distance of $r_{200}$ from the cluster centre, the same set of particles upon which the 3D fit is based. We will refer to this method as ``spherical WL''. One problem with this approach is that the expression of \citet{Wright_Brainerd_2000} for $g$ assumes a matter distribution extending to infinity, so fitting it to a catalogue of galaxy distortions based only on the matter distribution inside $r_{200}$ alone, which necessarily contains less total mass, will result in severe biases in both mass and concentration\footnote{The latter because the influence of the `missing mass' is greatest at large projected radii $R$ where only a short section of the line of sight passes through the inner cluster region as illustrated in Fig.~\ref{cutofftest} in appendix \ref{app1}.}.  We overcome this by instead fitting the reduced shear from an NFW profile that is, like our data, truncated at $r_{200}$ (\citealt{Takada_Jain_2003a}; \citealt{Takada_Jain_2003b}). We note that it is straightforward to also fit the projected mass profile of \citet{Takada_Jain_2003a} directly to the projected matter within $r_{200}$. This method approaches the 3D fit even closer, with the only remaining difference being a fit in 2D vs.~one in 3D. We have done this, and found that the results are in very close agreement with those of the `spherical WL' method.

\begin{figure}
\includegraphics[width=\columnwidth]{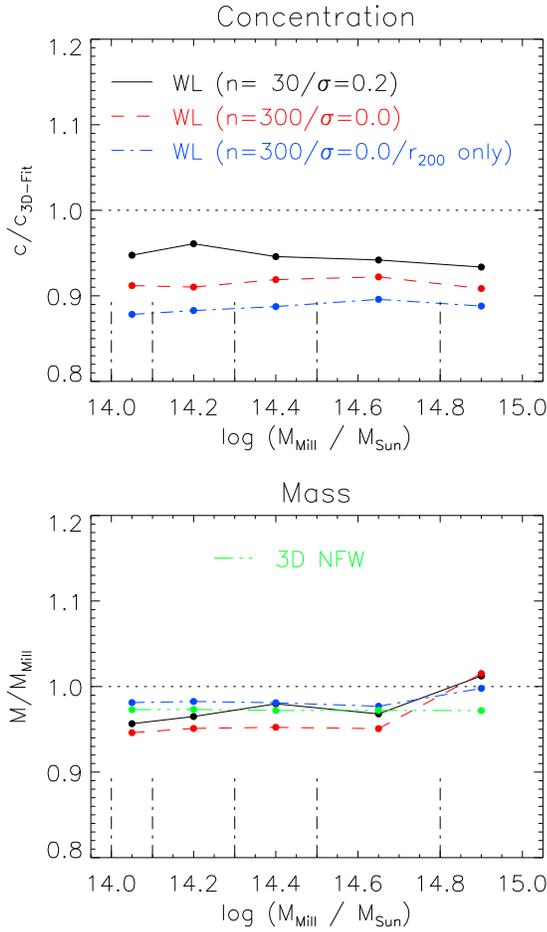}
\caption{Median values of the distributions derived from our default WL simulation ($n = 30, \sigma = 0.2$, black solid line), ``perfect WL'' ($n = 300, \sigma = 0.0$, red dashed line) and ``spherical WL'' (parameters as in perfect WL, but only based on the matter distribution within $r_{200}$, blue dash-dot line). See text for details. \emph{Top:} concentration, \emph{bottom:} mass, also showing the median values in $\mdd$ obtained from 3D mass profile fitting (green dash-dot-dot-dot line). The vertical lines at the bottom show the underlying mass bins.}
\label{medians}
\end{figure}

\begin{figure}
\includegraphics[width=\columnwidth]{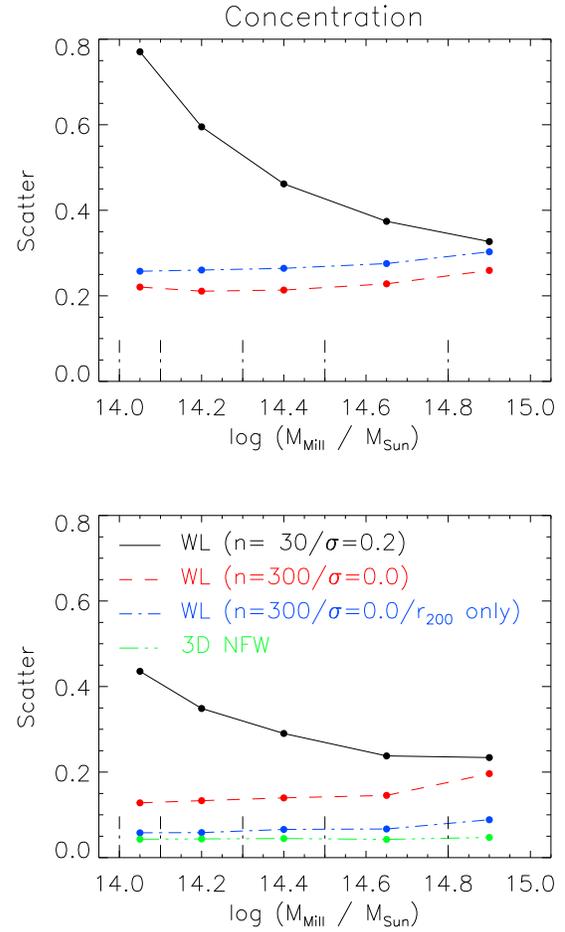}
\caption{As Fig.~\ref{medians}, but instead showing the scatter $e^\sigma-1$ in the best-fit lognormal distribution with $\sigma$ as defined in equation \eqref{lognormalfit}.}
\label{fig:scatter}
\end{figure}

The bias and scatter as defined in equation \eqref{lognormalfit} in masses and concentrations resulting from both these methods are shown in Fig.~\ref{medians} and \ref{fig:scatter} respectively, the perfect WL fit represented by red lines, the spherical WL fit by blue ones. For ease of comparison, we also include the values determined from our default WL simulation as found in Table \ref{table} as black lines. 

\subsection{Influence of background galaxies}
We can judge the influence of errors associated with the background galaxies by comparing the ``default'' and ``perfect'' WL reconstructions (black and red curves in Figs.~\ref{medians} and \ref{fig:scatter}), the latter one, as explained above, not being affected by them. 

\subsubsection{Scatter due to background galaxies}
Focusing first on scatter in default vs.~perfect WL (Fig.~\ref{fig:scatter}), we find a very similar picture for both mass and concentration: it is comparable for the highest-mass clusters, but while the latter is almost mass-independent, the former increases considerably with decreasing cluster mass. This is what would be expected from the influence of shape noise: Less massive haloes, producing a weaker shear signal, yield a lower signal-to-noise ratio than their high-mass counterparts; in the total absence of shape noise, however, the decreasing shear signal is irrelevant. 

\subsubsection{Bias due to background galaxies}
\label{sec:galbias}
As with scatter, the variation in bias between default and perfect WL is similar for both mass and concentration. In both cases, it is generally negative and slightly stronger in the case of perfect WL than in the default simulation (with the exception of the mass bias for the highest mass clusters): $\approx -9\%$ vs.~$\approx -5\%$ for concentration and $\approx -5\%$ vs.~$\approx -3\%$ for mass, with no clear trend with halo mass in any of them. This might be slightly surprising, given that we would not expect a direct bias from the unknown background galaxy ellipticities. However, in Section \ref{sec:biasorigin} below, we show that substructure tends to produce an overall negative bias. One possible explanation of the bias due to background galaxies is therefore that the combination of shape noise and finite sampling essentially smoothes out small-scale effects like substructure. This indirect bias can, overall, make imperfect weak lensing observations \emph{of very large cluster samples} slightly less biased than perfect ones. 

\subsection{Structure outside $r_{200}$}

We now look at the structure of the clusters themselves, focusing first on their outer regions beyond $r_{200}$ by comparing the ``perfect'' and ``spherical'' WL reconstructions (red and blue curves in Figs.~\ref{medians} and \ref{fig:scatter}) --- the only difference between them being that the former includes mass beyond $r_{200}$ while the latter does not. 

\subsubsection{Concentration}
For the concentrations, we find a similar scatter from both methods (compare the red and blue curves in the top panel of Fig.~\ref{fig:scatter}). In fact, the scatter is even slightly larger for the spherical analysis, a strong indication that it is driven by the mass distribution inside $r_{200}$ as discussed in detail below. The bias is even stronger for spherical WL than either of the other two methods (compare the blue and black/red curves in the top panel of Fig.~\ref{medians}), at a level of $\approx -12\%$ largely independent of cluster mass. This implies that effects {\it inside} $r_{200}$ bias the concentration low, whereas those \emph{outside} tend to increase it again, by $\approx 3\%$. The overall WL concentrations are biased negative, because the influence of the former effects dominates. One such effect causing a positive bias could be the deviation of the azimuthally averaged mass distribution from an NFW profile as recently shown by \citet{Oguri_Hamana_2011}.

\subsubsection{Mass}
\label{sec:mass}
Looking at the masses, a somewhat different picture emerges: The values reconstructed from the spherical fit (blue lines) are considerably more tightly constrained with a typical scatter of only $\sim 5\%$ and a bias at a level of only $\approx -2$\%. This becomes even more remarkable when compared to the spread in the masses obtained from 3D fitting, ie. $M_\text{3D} / M_\text{Mill}$, which we show by green lines in Figs.~\ref{medians} and \ref{fig:scatter}. The very close agreement between these indicates that the spherical WL mass reconstruction is as accurate as could be hoped for. The negative bias in the `perfect' and `default' simulations is thus due to the mass distribution outside $r_{200}$. As in the case of concentration, a likely explanation for this bias is the deviation of the mass distribution outside $\sim r_{200}$ from an NFW profile (see, e.g., Fig.\ 1 of \citealt{Hayashi_White_2008}). This leads to less mass along the line of sight than expected from an NFW profile extending to infinity, which explains the negative mass bias (see also \citealt{Oguri_Hamana_2011}).   

\subsection{Deviations from NFW within $r_{200}$: Triaxiality and substructure}

We demonstrated above that most of the scatter and bias in reconstructed concentrations is due to deviations from a spherically symmetric NFW profile within $r_{200}$. The obvious culprits responsible for these deviations are asphericity (e.g., triaxial haloes) and substructure. We now investigate the role played by each of these sources of error.

To begin with, an overall sense of the validity of the (spherically averaged) NFW approximation is given by the $M_\text{3D} / M_\text{Mill}$ distributions shown in Fig.~\ref{medians} and \ref{fig:scatter} (green lines). With $\mmill$ being a model-independent quantity, it provides a reference value for $\mdd$ which can be used as an indication of how well the NFW profile describes a cluster. The (logarithmic) scatter and bias are both small at a level of $\approx 5\%$ and $\approx -2\%$ respectively. This confirms that, despite the obvious presence of substructure and the overall triaxiality of the simulated clusters (see, e.g., Fig.~\ref{projections}), the \emph{spherically averaged} density is still well-described by the NFW profile. 

But while the 3D mass structure of the haloes may be relatively well described by an NFW profile for many clusters, the lensing deflection depends on (the gradient of) the \emph{projected} density. For realistic, non-spherically-symmetric cluster haloes, deviations from symmetry can be expected to affect 3D and 2D reconstructions differently, so it is entirely plausible that the shear signal for a cluster, even one that is well represented by an NFW profile in 3D, may not be well described by a 2D profile derived from it.  

\subsubsection{Halo triaxiality}
We first look at the effect of halo triaxiality. \citet{Jing_Suto_2002} suggest a triaxial generalisation of the spherical NFW profile by replacing the radius $r$ in \eqref{nfwrho} by $\reff$ where
\begin{equation}
\reff^2 = \frac{X^2}{a^2}+Y^2+\frac{Z^2}{b^2}
\end{equation}
where $X$, $Y$ and $Z$ are the distances along the major, intermediate and minor axes respectively; the numbers $a$ and $b$ are the ratio of the major and minor axis to the intermediate axis, respectively.

In a similar way to our standard 3D fitting procedure described above, we now also fit this triaxial NFW profile to our lensing haloes. Each halo is first subdivided into five concentric shells covering a radial range from -1.5 to 0 in $\log_{10} (r/r_{200})$. Each shell is further divided into $24^2$ sectors of equal volume. Overall, this procedure divides the cluster into 2880 cells, for each of which we compute the average density $\rho_i$. The triaxial NFW profile is then fit by least-squares regression\footnote{We have also tried computing the axis ratios by diagonalising the moment-of-inertia tensor (see e.g., \citealt{Shaw_et_al_2006}).  In general the two methods produce results in the same sense (i.e., both favour ``prolate-like'' distributions), with the moment-of-inertia method generally yielding slightly smaller axis ratios than found from our default method.}.

\begin{figure*}
\includegraphics[width=180mm]{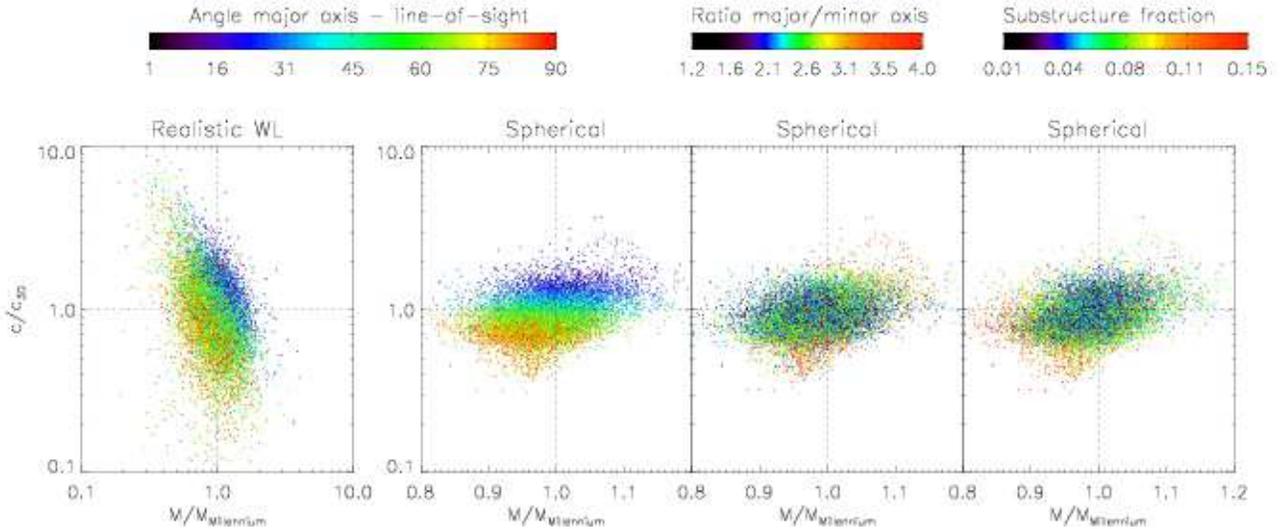}
\caption{Reconstructed masses and concentrations, both normalised to ``true'' values ($\mmill$ and $\cdd$ respectively). The values in the left-most panel are derived from our default WL analysis, in the other three from the spherical WL reconstruction based only on the matter distribution inside $r_{200}$. Note that the horizontal axes are scaled differently to account for the much smaller mass scatter in the latter reconstruction. The colour shows, from left to right, the angle between the major halo axis and line of sight (\emph{first two panels}), the ratio between major and minor halo axis (\emph{third panel}) and the fraction of the halo mass in bound substructures within $r_{200}$ (\emph{last panel}). The influence of halo orientation is clearly dominating the scatter.}
\label{fig:mcrelative}
\end{figure*}

Applying this to the question of the influence of halo triaxiality on lensing reconstructions, we show, in the first panel of Fig.~\ref{fig:mcrelative}, how the angle $\delta$ between the line of sight and the major halo axis correlates with over- and underprediction of the lensing cluster's mass and concentration. Each point corresponds to a mock WL reconstruction of a simulated cluster; we include all simulated clusters with 5 projections each. It is immediately obvious that projections with small $\delta$ lead to overpredictions in both mass and concentration, the opposite being true for cases with $\delta \sim 90^o$. The role of orientation becomes even more obvious when we move to the spherical WL fit described above and only analyse the region inside $r_{200}$, as shown in the second panel in Fig.~\ref{fig:mcrelative}. This reduces the scatter along the (1,-1) direction considerably, identifying the influence of background galaxies on low-mass clusters as its main cause. The finding that decreasing $\delta$ increases $\cwl$ and $\mwl$ \emph{of individual clusters} simultaneously confirms previous work on the effect of cluster triaxiality by \citet{Clowe_et_al_2004}, who analysed a sample of four massive haloes and found a strong correlation between concentration and halo orientation, and \citet{Corless_King_2007} who derived a similar result using analytic cluster haloes. The latter authors concluded that triaxiality could cause overpredictions in concentration by a factor of 2, in good agreement with the second panel of Fig.~\ref{fig:mcrelative}. The scatter in masses --- deviations up to a factor of 1.5 --- reported by these authors, on the other hand, is considerably larger than ours ($\sim 1.1$). This may be due to the best-fit NFW mass being influenced most severely by matter in the cluster outskirts as discussed above. The analytic model of \citet{Corless_King_2007} includes this matter beyond $r_{200}$ whereas our spherical WL analysis does not.

Note that in the first two panels of Fig.~\ref{fig:mcrelative} the \emph{magnitude} of the ratio between major and minor axis was not taken into account at all. We explore its influence in the third panel of Fig.~\ref{fig:mcrelative} and find a much weaker correlation than with halo orientation. Haloes with extremely high axis ratios have a tendency to lead to more over- or underpredicted masses and concentrations, as should be expected, but the influence is clearly much smaller than that of orientation. 

We noted above that the concentration scatter \emph{decreases} slightly upon inclusion of matter outside $r_{200}$. One possible explanation for this effect is that matter beyond $r_{200}$, while still correlated to some extent with the cluster major axis, is more randomly distributed (see also \citealt{Becker_Kravtsov_2011}) and therefore reduces the impact of the triaxiality-induced concentration scatter a bit.

\subsubsection{Substructure}

While we have just demonstrated a strong influence of halo triaxiality on reconstructed NFW parameters \emph{of individual clusters}, there remains some scatter which is not correlated with this. In the right-most panel of Fig.~\ref{fig:mcrelative} we therefore investigate the influence of substructure, quantified as the fraction $\fsub$ of the halo mass which is found in bound subhaloes within $r_{200}$ identified by the \textsc{Subfind} algorithm \citep{Springel_et_al_2001}. This fraction is generally small with only 15\% of haloes having $\fsub > 0.1$. The plot shows some influence on the reconstructions, most notably on concentration, which tends to be underpredicted in high-substructure clusters. However, as with the axis ratio, the influence is much less strong than that of cluster orientation.

\subsubsection{Origin of concentration bias}
\label{sec:biasorigin}
\begin{figure}
\includegraphics[width=80mm]{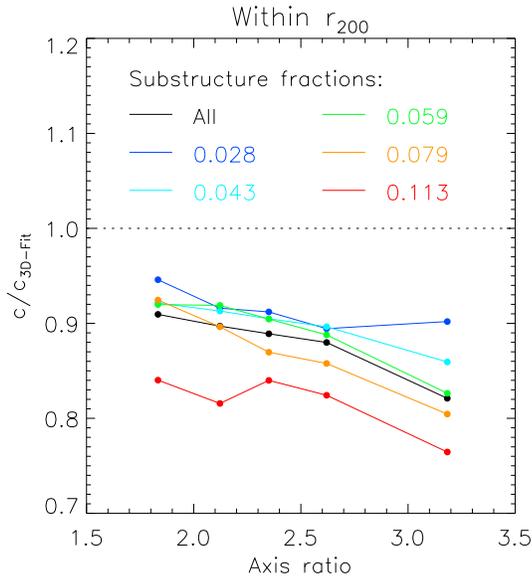}
\caption{Influence of substructure fraction (different colour lines) and triaxiality (x-axis) on concentration bias (y-axis). It is evident that higher substructure and higher triaxiality both lead to lower concentrations, but the effect of substructure is stronger.} 
\label{fig:biasresolution}
\end{figure}

We have shown above that the scatter in concentration can be ascribed largely to the influence of halo orientation. However, for a large sample of clusters, oriented randomly towards the observer, this leads to overestimation of concentration in some haloes, and underestimation in others (see Fig.~\ref{fig:mcrelative}). While this does not necessarily eliminate \emph{all} influence of triaxiality on the concentration bias even for arbitrarily large cluster samples, it can be expected to reduced it to a level where the effect of substructure becomes non-negligible. We explore the role of both effects in this context in Fig.~\ref{fig:biasresolution}. For this, the sample is split into five equally large quintiles, once in order of increasing axis ratio, and once by substructure fraction. For each of the 25 resulting combinations, the median concentration, normalised to $\cdd$, is then formed. In this way, the figure shows both the variation with axis ratio at (nearly) fixed substructure level and vice versa and can therefore eliminate potential correlations between these two effects. For added clarity, the black line shows the concentration bias when the sample is split only according to axis ratio.

There are two clear trends in this figure: A decrease in concentration from left to right, corresponding to an increase in axis ratio (of $\approx 5\%$ in the case of the lowest substructure quintile), and from blue to red, as the substructure fraction increases (of $\sim 10\%$ for the least elongated clusters). Both triaxiality and substructure therefore tend to lower the concentration, but the influence of substructure appears to be dominant.   

While the addition of substructure to the outer parts of a smooth NFW halo can be expected to reduce the best-fit concentration of both 2D and 3D analyses, its influence appears to be larger in 2D. This might be due to the low surface mass density in the projected outer cluster regions (see Fig.~\ref{cutofftest}), which is affected more severely by the presence of substructure than the mass profile obtained in 3D.   

\subsection{Summary of this section}

To summarise the results and interpretations of this section: Within $r_{200}$, substructure and triaxiality cause a negative bias ($\sim -12\%$) and strong scatter in WL concentration, but leave the reconstructed mass largely unaffected. Structure outside $r_{200}$, the ``correlated large-scale structure'', biases concentration positively and mass negatively, due to deviations of the mass profile from the NFW model at these large radii. The scatter in mass is strongly increased by matter in this region, but the orientation-induced scatter in concentration is, to a small degree, cancelled out by it. Imperfect weak lensing observations smooth out part of the influence of substructure and therefore cause a slight improvement in the mass and concentration bias. Shape noise, while not solely responsible for the mass and concentration scatter, contributes considerably to it, in particular for lower mass clusters. 

\section{Discussion and Conclusions}
\label{sec:conclusion}

We have analysed a large set of mock WL observations of galaxy clusters at redshift $z_L \simeq 0.2$ extracted from the Millennium Simulation (MS) using background galaxies with elliptical isophotes at uniform redshift $z_S = 1.0$. The aim of our work was to quantify the expected scatter and bias in the reconstructions of masses and concentrations of large cluster samples derived from WL surveys such as DES and LSST.  We focus on the effect of the matter within and close to the lensing clusters, explicitly ignoring the effect of uncorrelated large-scale structure (as studied, e.g., by \citealt{Hoekstra_et_al_2011}, see below).  Furthermore, as we are interested in the bias and scatter induced by the clusters and the weak lensing method itself, we have explicitly ignored any additional observational sources of error, such as difficulties in measuring the background galaxy ellipticities and redshifts.  Nevertheless, from the analysis of these mock observations, and subsequent comparison with the true cluster parameters, we can draw the following conclusions:

\begin{itemize}
\item We confirm that the dark matter haloes in our sample, although not selected using any relaxation criteria, are well-described by the NFW profile. The NFW masses $M_\text{3D}$ agree to within $\approx 5 $ \% ($1\sigma$) with the true halo masses. 
\item The mass-concentration relation derived from our mock observations has a normalisation of $\alpha = (4.25 \pm 0.04)$ and slope $\beta = (-0.10 \pm 0.01)$. The corresponding values for the `true' mass-concentration relation of our sample are $\alpha_{3D} = (5.02 \pm 0.08)$ and $\beta_{3D} = (-0.16 \pm 0.02)$. Both the slope and normalisation of the weak lensing derived relation are too low compared to their 3D counterparts.

\item The spread in the WL parameters $\mwl/\mmill$ and $\cwl/\cdd$ closely follows a log-normal distribution. Its scatter decreases with increasing halo mass, due to an increased strength of the lensing signal (for concentration from 0.57 to 0.29, and for mass from 0.36 to 0.21 between clusters with $M_{200} < 10^{14.1} M_\odot$ and $M_{200} > 10^{14.8} M_\odot$ respectively). Both mass and concentration show in general a negative median bias at a level of $\approx -3\%$ and $\approx -5\%$ respectively. Due to the presence of scatter, the mean mass and concentration are generally higher than their median counterparts.

\item While shape noise due to the unknown intrinsic orientation of background galaxies is an important contribution to the parameter scatter (particularly for lower mass haloes with $M_{200} \sim 10^{14} M_\odot$), we find that physical properties of the clusters themselves (e.g., triaxiality, substructure, correlated line-of-sight matter) also contribute significantly and are the dominant source of scatter for haloes with masses of $M_{200} \sim 10^{15} M_\odot$. Triaxiality of the cluster halo within $r_{200}$, is the main contributor to the scatter in concentrations, alignment between major halo axis and the line-of sight being the dominant factor. The concentration bias is dominated by substructure within the cluster. The bias and scatter in the recovered mass is affected mostly by matter beyond $r_{200}$, i.e., correlated large-scale structure where deviations from the NFW profile may be important.  
\end{itemize} 

We note that this comparison between WL-derived and `true' 3D-derived concentrations is somewhat sensitive to the radial fitting range used in determining the latter; as found by \citet{Gao_et_al_2008}, a larger fitting range probing regions closer to the cluster centre tends to increase the best-fit concentration. In this work, we have used the smallest possible truncation radius allowed by the resolution of the simulation and the mass range of our clusters ($0.02\,r_{200}$). For an accurate calibration of the bias and scatter expected from a particular survey, it is therefore not only important to accurately model the survey, but also the way in which the theoretical comparison data is derived. At the price of introducing a third fitting parameter, use of the Einasto profile \citep{Einasto_1965} may be a possible way to remove this dependence on the radial fitting range.

Our analysis of the physical origin of scatter and bias suggests that the most promising way to reduce the uncertainty in concentration measurements from WL is to employ triaxial halo models as suggested by \citet{Corless_King_2008}. It is however important to keep in mind that not only the analysis of observational data, but also the theoretical predictions used for comparison need to use this triaxial model because the halo `mass' and `concentration' in a triaxial model will in general be different from their counterparts derived using spherical averaging.

As stated above, in the present study we have not included the influence of uncorrelated large-scale structure at large distances from the cluster. Using spherical analytic NFW haloes, \citet{Hoekstra_et_al_2011} have recently studied its effect on the recovered masses and concentrations and their correlation.  They found a small bias in the \emph{slope} of the recovered mass-concentration relation, in the sense that the recovered relation is slightly steeper than the true underlying one. Additionally, their study addresses the influence of uncorrelated large-scale structure on the scatter in both mass and concentration: for a cluster of mass $M_{200} = 10^{15} M_\odot$ this is comparable to the effect of shape noise and can therefore be expected to contribute significantly to the overall error budget, in particular for surveys extending to lower-mass clusters. A combined study, taking into account both uncorrelated and correlated error sources using realistic clusters would be a valuable way to make even more realistic predictions of the expected bias and scatter in WL parameter reconstructions.   

Another potentially important factor that should be explored in future realistic mock weak lensing surveys is that of realistic cluster selection.  In the present study, we have imposed a simple (true) halo mass cut to establish first whether or not the recovered masses and concentrations are biased, which they are, for an underlying sample that is unbiased.  Cluster selection itself (e.g., based on optical richness, Sunyaev-Zel'dovich effect flux, X-ray luminosity, weak lensing shear signal), however, can potentially introduce biases which may be larger than those we have studied here and therefore it is important that these potential biases be quantified in the future.

While our quantification of scatter and bias in weak lensing measurements was motivated by the use of the mass-concentration relationship to constrain cosmological parameters, our results can also be applied to outliers in observed concentration. For instance, we have shown that those clusters which are elongated along the line of sight are most liable to having their concentration overestimated in weak lensing studies.  We can also make predictions for \emph{how many} clusters we would expect to exceed a particular concentration. For high-mass system with $M_{200} > 10^{14.8} M_\odot$, for which the Millennium Simulation includes no systems above $\cdd = 5.7$, we find that 7\% of cluster projections yield weak lensing-derived concentrations of $\cwl > 6$, and only 1\% exceed $\cwl = 8$. We re-iterate at this point that our study has not taken into account uncorrelated line-of-sight structure, which might increase these fractions somewhat.    

As long as bias and scatter are properly accounted for, future WL surveys are expected to accurately determine the mass-concentration relationship of galaxy clusters, even with modest numbers of background galaxies per cluster.  The form of the mass-concentration relationship, and its evolution with redshift, provide a sensitive probe of the growth of structure and therefore offer an important new and independent method of testing our cosmological paradigm.

\section*{acknowledgements}
The authors thank the anonymous referee for their many constructive suggestions which significantly improved the paper.
The authors thank the Virgo Consortium for providing the Millennium Simulation particle data and Simon White for helpful discussions.  YMB acknowledges a postgraduate award from STFC. IGM is supported by a Kavli Institute Fellowship at the University of Cambridge. LJK is supported by a University Research Fellowship from the Royal Society. This research has made use of the \textsc{Darwin} High Performance Computing Facility at the University of Cambridge. 

\bibliographystyle{mn2e}
\bibliography{Lensing}

\appendix
\section{Truncated NFW profile}
\label{app1}
The NFW profile truncated at some limiting (3D) radius $\rlim$ (\citealt{Takada_Jain_2003a}, \citealt{Takada_Jain_2003b}) is
\begin{equation}
\rho = \frac{\rho_s}{(r/r_s)(1+r/r_s)^2}\Theta(\rlim - r)
\end{equation}
where $\Theta$ is the Heaviside Unit Step function and $\rho_s$ and $r_s$ the scale density and length as defined in section \ref{sec:fitting}. 
 
The projected mass density of this profile integrated along the line of sight is then given by
\begin{equation}
\Sigma(R) = 2\rho_s r_s \int_R^{\rlim}\frac{1}{\sqrt{r^2-R^2}(1+r/r_s)^2} dr 
\end{equation}
where $R$ is the distance from the cluster centre perpendicular to the line of sight.
 
Evaluating this integral yields
\begin{equation}
\Sigma = 2\delta_c \rho_c r_s S
\end{equation}
where S depends on the value of $x = r/r_s$.

For $x < 1$:
\begin{equation}
S = \frac{\ln\left(\frac{x(r_s+\rlim)}{\rlim+r_s x^2 - \sqrt{(1-x^2)(\rlim^2-r_s^2x^2)}}\right)-\frac{\sqrt{(1-x^2)(\rlim^2-r_s^2x^2)}}{r_s+\rlim}}{\left(1-x^2\right)^{3/2}}
\end{equation}

For $x = 1$:
\begin{equation}
S = \frac{(\rlim-r_s)(2r_s+\rlim)}{3(r_s+\rlim)\sqrt{\rlim^2-r_s^2}}
\end{equation}

For $x > 1$:
\begin{equation}
S = \frac{-\arctan\left(\frac{\sqrt{(\rlim^2-r_s^2x^2)(x^2-1)}}{\rlim+r_sx^2}\right)+\frac{\sqrt{(x^2-1)(\rlim^2-r_s^2x^2)}}{r_s+\rlim}}{(x^2-1)^{3/2}}
\end{equation}

\begin{figure}
\includegraphics[width=\columnwidth]{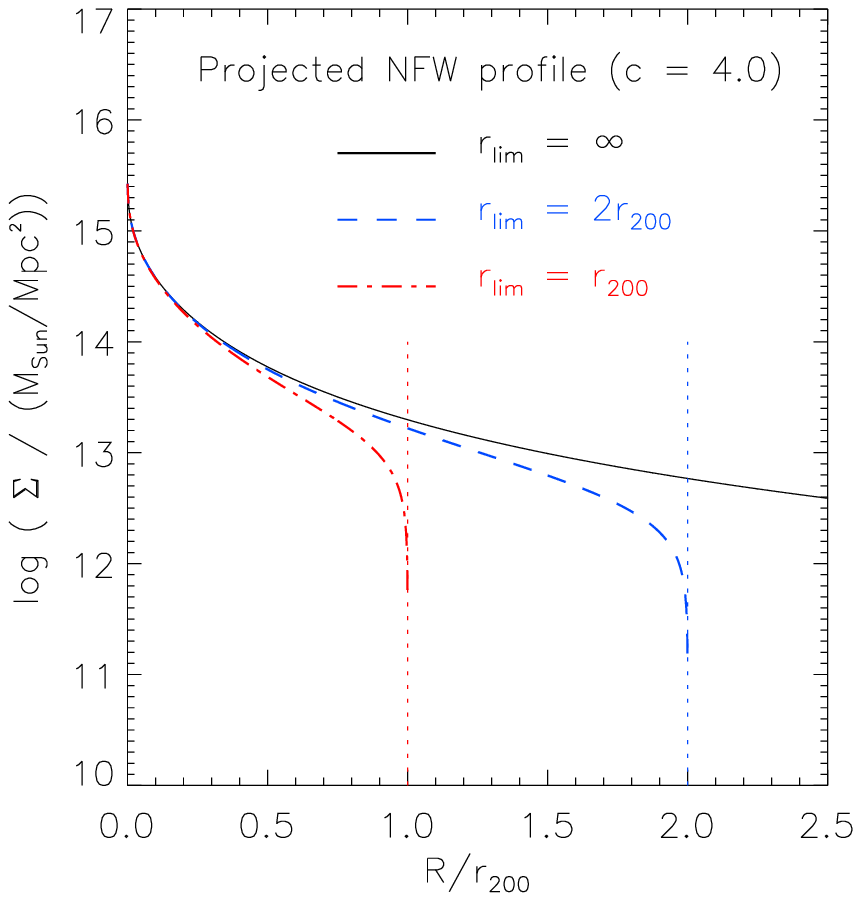}
\caption{Surface density of an NFW profile integrated along the line of sight. The solid black line corresponds to an untruncated infinite profile, whereas the blue dashed and red dash-dot lines show profiles truncated at $2\,r_{200}$ and $r_{200}$ respectively. Truncation mostly affects the surface density in the outer cluster regions beyond $\sim 0.5 \rlim$.} 
\label{cutofftest}
\end{figure}

\section{Effect of variations in lensing simulation}
\label{sec:appb}
\subsection{Line-of-sight integration length}
\label{sec:testlength}
\begin{figure*}
\includegraphics[width=150mm]{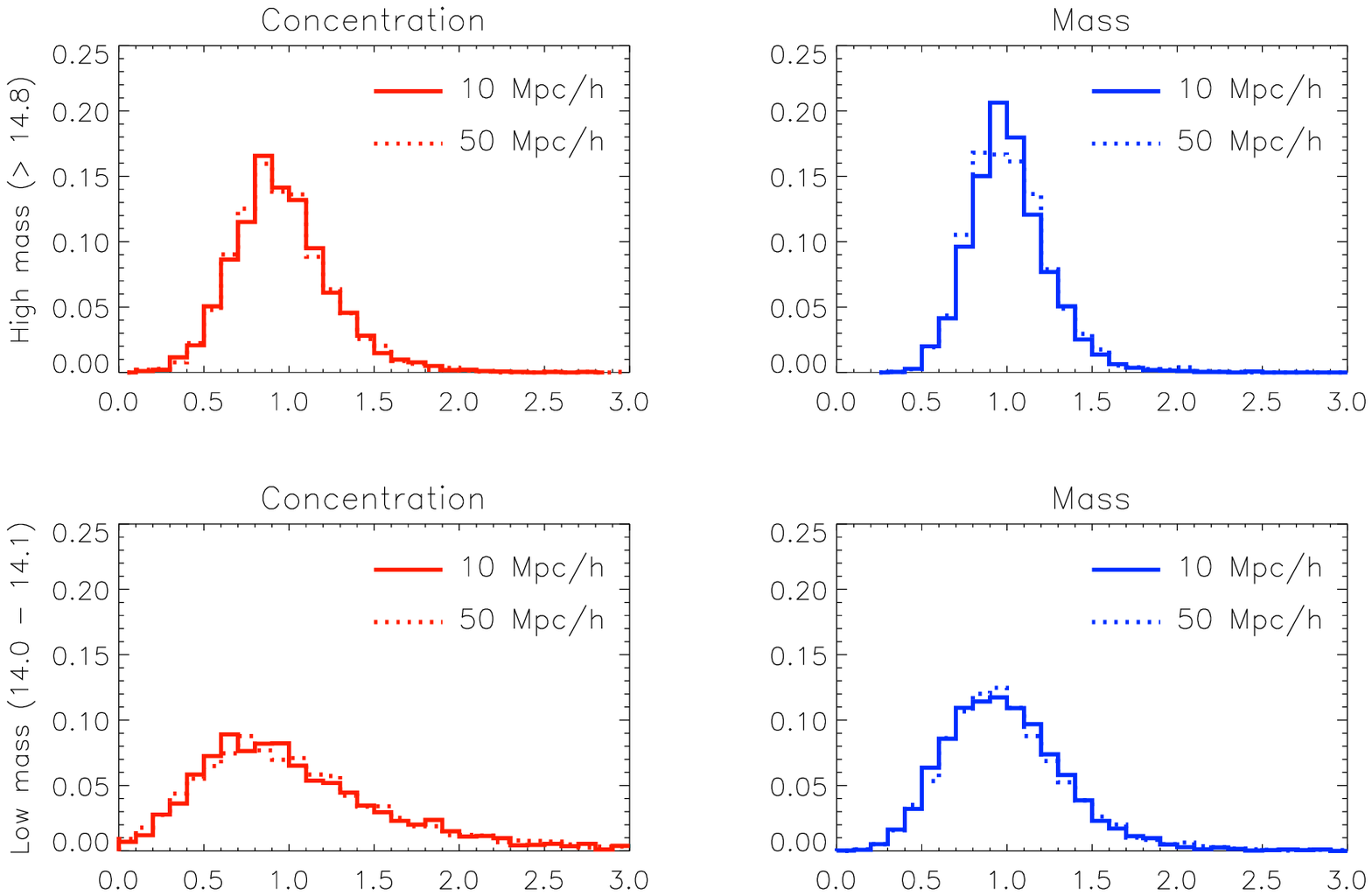}
\caption{Effect of increasing the line-of-sight integration length to 50 $h^{-1}$ Mpc, shown by solid lines, compared to our default choice of 10 $h^{-1}$ Mpc (broken lines) on concentrations (\emph{left}) and masses (\emph{right}). There is very little variation, neither for high-mass systems with $M_{200} > 10^{14.8} M_\odot$ (\emph{top}), nor for low-mass systems with $M_{200} < 10^{14.1} M_\odot$ (\emph{bottom}).}
\label{fig:boxsizetest}
\end{figure*}

The effect of increasing the line-of-sight integration length from our default value of $10 h^{-1}$ Mpc by a factor of 5 to $50 h^{-1}$ Mpc is explored in Fig.~\ref{fig:boxsizetest}. We find very little variation between the two lengths, consistent with the findings by \citet{Becker_Kravtsov_2011}. Our default integration length of $10 h^{-1}$ Mpc is therefore sufficiently large to capture the influence of correlated large-scale structure close to the lensing cluster.

\subsection{Background galaxy ellipticity}
\label{sec:sigmatest}
\begin{figure*}
\includegraphics[width=150mm]{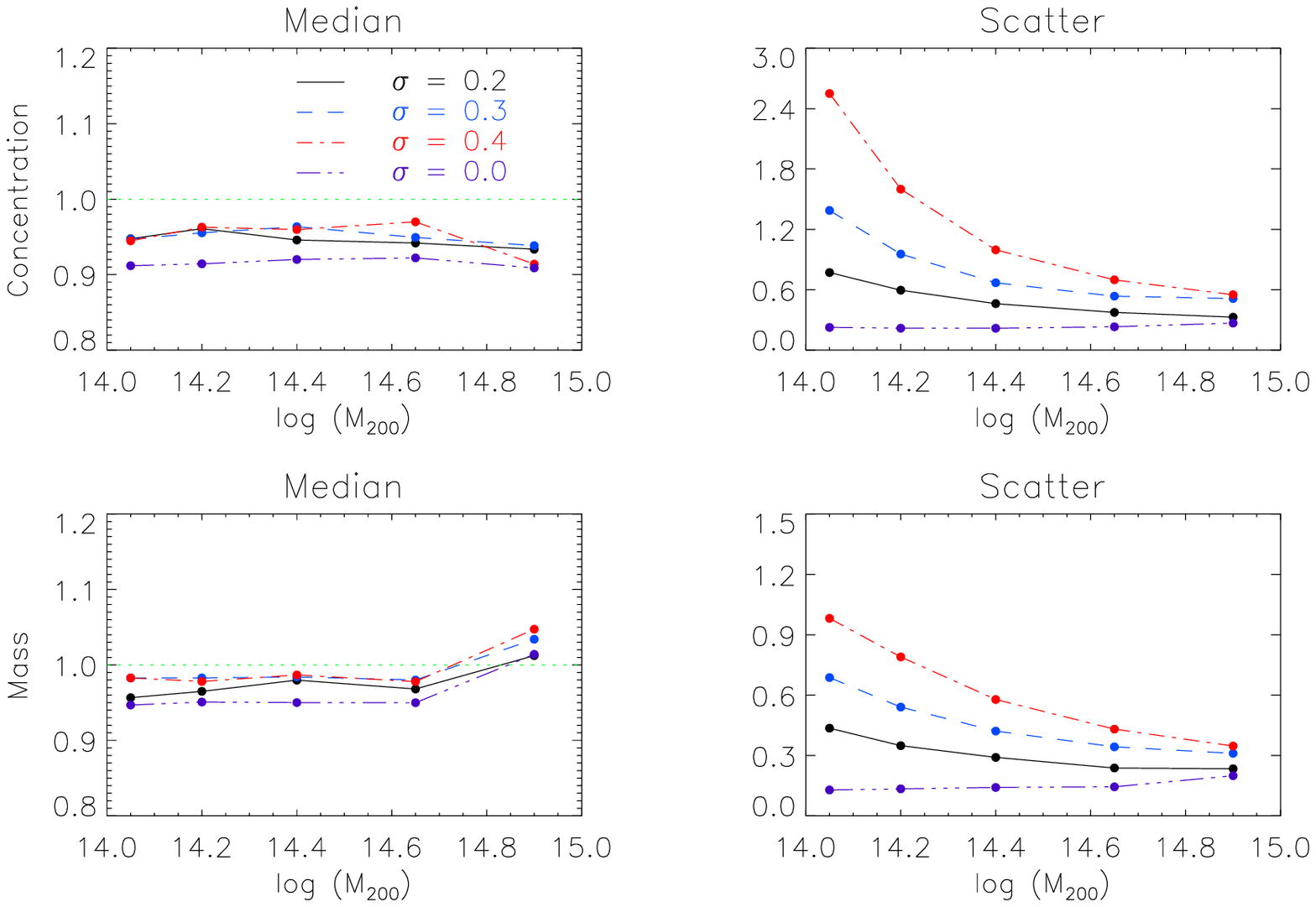}
\caption{Effect of increasing the dispersion in the intrinsic ellipticities of the background galaxies from our default value of 0.2 per component (black, solid line) to 0.3 (red, dashed line) and 0.4 (blue, dot-dashed line) per component respectively. The top two panels show the effect on the concentrations, the bottom two on the masses. In both cases medians are shown on the left and scatter on the right; note the different ordinate scalings for the mass and concentration scatter plots. Increased shape noise causes a small positive bias, due to smoothing out of substructure and triaxiality, and a strong increase in scatter.}
\label{fig:sigmatest}
\end{figure*}

In Fig.~\ref{fig:sigmatest} we investigate the effect of increasing the value of the background galaxy ellipticity dispersion from our default of 0.2 to 0.3 and 0.4 per component as well as decreasing it to zero. As can be expected, the scatter in both mass and concentration increases with increased shape noise. There is also a very small influence on the mass and concentration bias, in the sense that a higher ellipticity dispersion leads to higher bias. As discussed in section \ref{sec:galbias}, this is most likely due to the fact that shape noise, in combination with finite sampling, smoothes out the influence of triaxiality and substructure, which both tend to cause a negative bias.  

\subsection{Weak lensing survey range}
\label{sec:cuttest}
\begin{figure*}
\includegraphics[width=160mm]{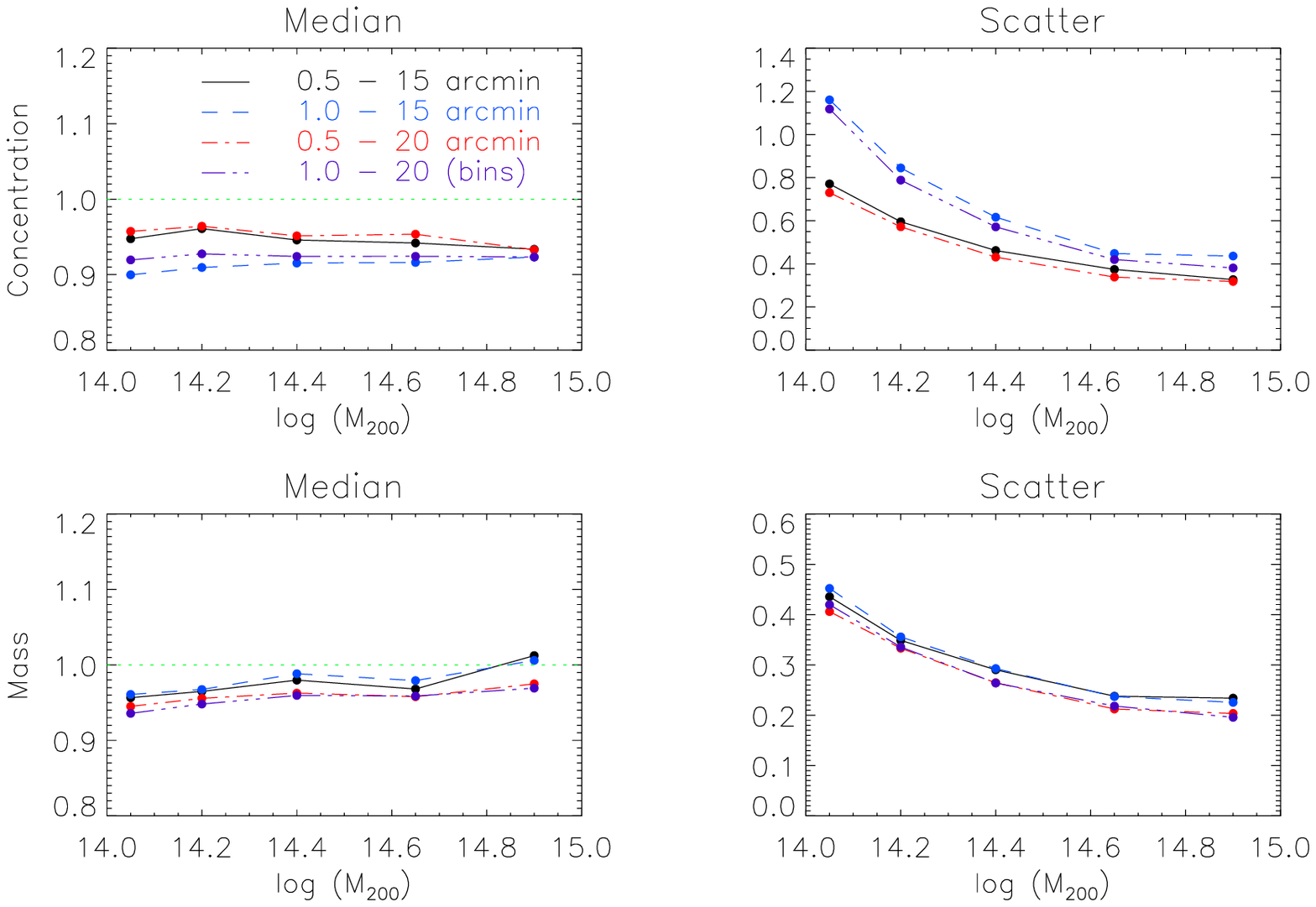}
\caption{Effect of varying the inner and outer cutoff radius ($r_\text{in}$ and $r_\text{out}$ respectively); note the different ordinate scalings for the mass and concentration scatter plots. Varying $r_\text{in}$ influences mostly the concentrations, whereas the masses are more sensitive to the value of $r_\text{out}$. For the first three analyses (black solid, red dashed and blue dash-dot lines), we use our default method of using each galaxy ellipticity individually, whereas the last one (purple dash-dot-dot-dot lines, 1$^\prime$ - 20$^\prime$) uses 15 logarithmically spaced bins to reproduce the analysis setup of BK11. There is no indication for significant differences between these two methods.}
\label{fig:cuttest}
\end{figure*}

Finally, we show in Fig.~\ref{fig:cuttest} the effect of varying the radial survey range from our default of $30^{\prime\prime}$ to $15^\prime$. The biases and mass scatter are only affected at a level of a few percent. The concentration scatter is somewhat more sensitive, particularly on the choice of inner cut-off radius: increasing this to $1^\prime$ causes a scatter increase by $\approx 30\%$ for the lowest-mass clusters. In general, the concentration appears to be more sensitive to the value of the inner than the outer cut-off radius, the opposite being true for mass. 
For ease of comparison with \citet{Becker_Kravtsov_2011}, we also include an analysis covering the radial range used in their study, in which we form a shear profile from 15 bins spaced logarithmically between $1^\prime$ and $20^\prime$ which is then fit to the NFW model. There is hardly any difference between the results from this fit and our default method in which we use each tangential galaxy ellipticity individually.

\section{Variation in bias definition}
\label{sec:mean}
\begin{figure}
\includegraphics[width=\columnwidth]{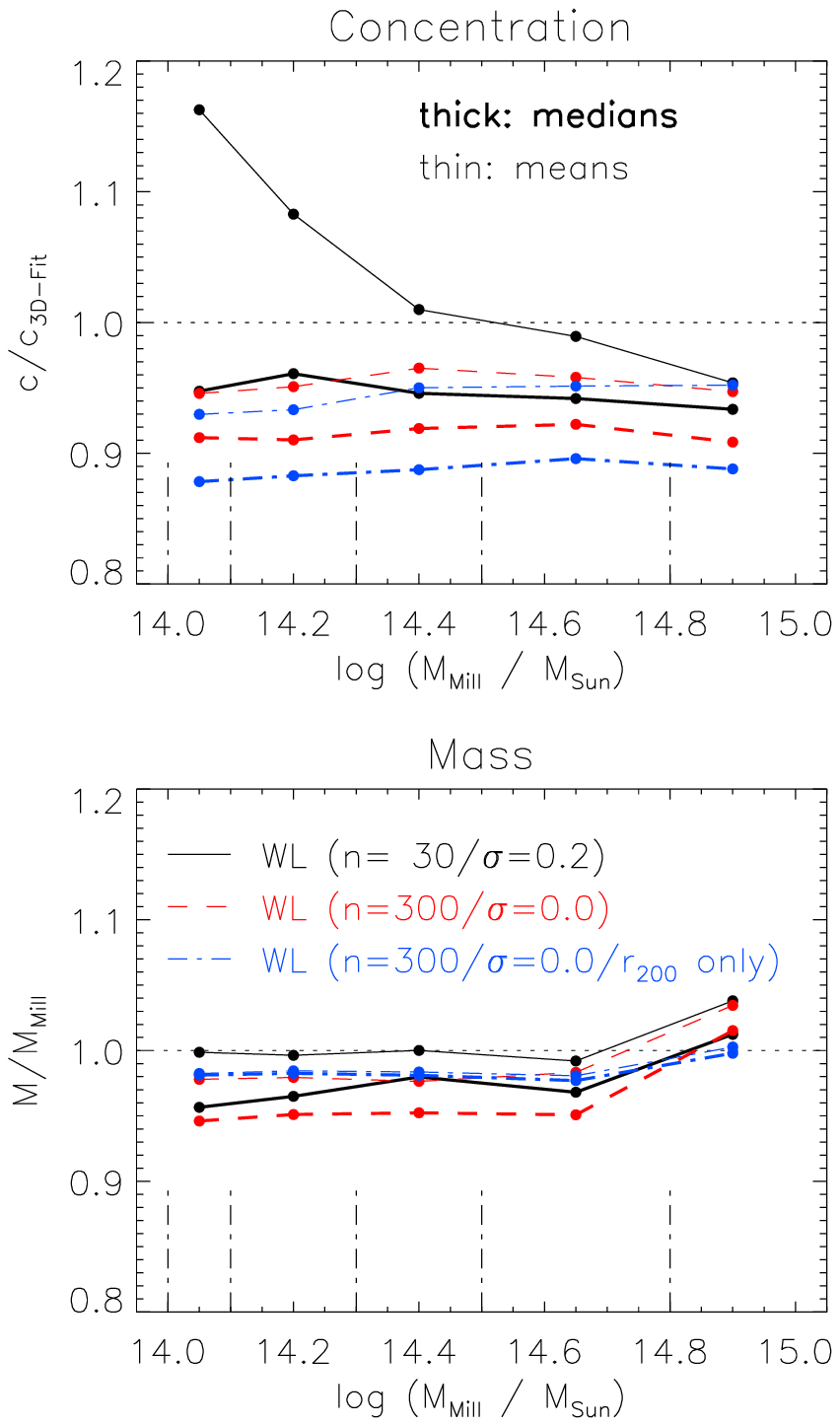}
\caption{Difference between bias definition as median (our default, thick lines) or mean (thin lines) on concentration (\emph{top}) and mass (\emph{bottom}). The mean shows, in general, a clear positive bias.}
\label{fig:mean}
\end{figure}

In Fig.~\ref{fig:mean}, the difference between defining bias as the median or mean is shown. While both are identical in the case of a Gaussian (normal) distribution, this is not generally the case for a log-normal parameterisation as employed here (see Fig.~\ref{histogram}). In terms of the parameters $\mu$ and $\sigma$ as defined in equation \eqref{lognormalfit}, the median is given by 
\begin{equation}
\label{eq:median}
\tilde{x} = e^{\mu}
\end{equation}
and the mean by 
\begin{equation}
\label{eq:mean}
<x>\, = \, e^{\mu + \sigma^2/2}.
\end{equation}
Thus, while the median depends only on $\mu$, the mean is additionally sensitive to the scatter $\sigma$. This is confirmed by Fig.~\ref{fig:mean}: Distributions with large scatter like concentration in our default analysis (black lines, top panel), have a mean up to $\approx 20\%$ higher than the median. For the very well constrained spherical WL mass distribution (blue lines, lower panel), on the other hand, mean and median are virtually identical. 

\end{document}